\documentstyle[amstex,12pt]{article}
\begin{document}

\author{{\bf Fabio Cardone}$^{a,b}${\bf , Alessio Marrani}$^{c}${\bf \ and } \and
{\bf Roberto Mignani}$^{b-d}$ \\
%EndAName
$a$ Istituto per lo Studio dei Materiali Nanostrutturati\\
(ISMN-CNR)\\
Via dei Taurini, 19\\
00185 ROMA, Italy\\
$b$ I.N.D.A.M. - G.N.F.M.\\
$c$ Dipartimento di Fisica ''E. Amaldi''\\
Universit\`{a} degli Studi ''Roma Tre''\\
Via della Vasca Navale, 84\\
00146 ROMA, Italy\\
$d$ I.N.F.N. - Sezione di Roma III}
\title{{\bf Killing symmetries of generalized Minkowski spaces.}\\
{\bf 3- Space-time translations in four dimensions}}
\maketitle

\begin{abstract}
In this paper, we continue the study of the Killing symmetries of a
N-dimensional generalized Minkowski space, i.e. a space endowed with a (in
general non-diagonal) metric tensor, whose coefficients do depend on a set
of non-metrical coodinates. We discuss here the translations in such spaces,
by confining ourselves (without loss of generality) to the four-dimensional
case. In particular, the results obtained are specialized to the case of a
''deformed'' Minkowski space $\widetilde{M_{4}}$ (i.e. a pseudoeuclidean
space with metric coefficients depending on energy).
\end{abstract}

\bigskip

\section{INTRODUCTION\protect\bigskip}

This is the third of a series of papers aimed at studying the Killing
symmetries of a $N$-dimensional generalized\bigskip\ Minkowski space, i.e. a
space endowed with a (in general non-diagonal)\ \bigskip metric tensor,
whose coefficients do depend on a set of non-metrical coordinates.\bigskip\
In the previous papers [1] and [2], we discussed both the
infinitesimal-algebraic and the finite-group structure of the
space-time\bigskip\ rotations in such a space.\

An example of a generalized Minkowski space is provided by the deformed
space-time $\widetilde{M_{4}}$ of {\em Deformed Special Relativity} (DSR).
DSR is a generalization of the {\em Standard Special Relativity} (SR) based
on a\ ''deformation'' of space-time, assumed\bigskip\ to be endowed with a
metric whose coefficients depend on the energy of\bigskip\ the process
considered [3]. Such a formalism applies in principle to {\em all} four
interactions (electromagnetic, weak, strong and\bigskip\ gravitational) ---
at least as far as their\bigskip\ nonlocal behavior and\ nonpotential part
is concerned --- and provides a metric representation of them (at\ least
for\bigskip\ the process and in the\ energy range considered) ([4]-[7]).
Moreover, it was shown\bigskip\ that DSR is actually a\ five-dimensional
scheme, in the\ sense that\bigskip\ the deformed Minkowski space can be
naturally embedded in\medskip\ a larger Riemannian manifold,\bigskip\ with
energy as fifth dimension [8].\bigskip\

In this paper, concluding the line of formal-mathematical research started
in [13] (followed by [12], [1] and [2]), we shall end our investigation by
discussing the space-time\bigskip\ translations in generalized Minkowski
spaces. For simplicity's sake, we shall restrict us (without loss of
generality) to the four-dimensional case, by specializing our results to the
deformed space-time $\widetilde{M_{4}}$ of DSR.

The organization of the paper is as follows. In Sect. 2 we briefly review
the\bigskip\ formalism of DSR \ and of the deformed Minkowski space $%
\widetilde{M_{4}}$. The results obtained in [1] and [2] concerning the
maximal Killing group of generalized \bigskip $N$-dimensional\ Minkowski
spaces are summarized in Sect. 3. Then, Sect. 4 gives a general treatment of
translation coordinate transformations in 4-d. generalized Minkowski spaces.
Sect. 5, concerning the Abelian Lie group $Tr.(3,1)_{DEF.}$ of deformed
space-time translations in DSR4, is divided in three Subsections: Subsect.
5.1 is about the 5-d. representation of infinitesimal contravariant
generators of $Tr.(3,1)_{DEF.}$; in Subsect. 5.2 the ''mixed'' deformed
Poincar\`{e} algebra is obtained, and 4-d. deformed Poincar\`{e} algebra
fully explicited; finally Subsect. 5.3 explicits the form of infinitesimal
and finite deformed translations in DSR4.

\section{\protect\bigskip DEFORMED\ SPECIAL RELATIVITY IN FOUR DIMENSIONS
(DSR4)}

The generalized (``deformed'') Minkowski space $\widetilde{M_{4}}$ (DMS4) is
defined as a\bigskip\ space with the same local coordinates $x$ of $M_{4}$
(the four-vectors of the usual\bigskip\ Minkowski space), but with metric
given by the metric tensor\footnote{%
In the following, we shall employ the notation ''ESC on'' (''ESC off'') to
mean that the Einstein sum convention on repeated indices is (is not) used.}%
\bigskip\
\begin{eqnarray}
g_{\mu \nu ,DSR4}(x^{5})
&=&diag(b_{0}^{2}(x^{5}),-b_{1}^{2}(x^{5}),-b_{2}^{2}(x^{5}),-b_{3}^{2}(x^{5}))=
\nonumber \\
&&  \nonumber \\
&&\stackrel{\text{{\footnotesize ESC off}}}{=}\delta _{\mu \nu }\left[
b_{0}^{2}(x^{5})\delta _{\mu 0}-b_{1}^{2}(x^{5})\delta _{\mu
1}-b_{2}^{2}(x^{5})\delta _{\mu 2}-b_{3}^{2}(x^{5})\delta _{\mu 3}\right] ,
\nonumber \\
&&
\end{eqnarray}
\bigskip where the $\left\{ b_{\mu }^{2}(x^{5})\right\} $ are dimensionless,
real, positive functions of the \bigskip independent, non-metrical (n.m.)
variable $x^{5}$ \footnote{%
Such a coordinate is to be interpreted as the energy (see Refs. [3]-[8]);
moreover, the index $5$ explicitly refers to the above-mentioned fact that
the deformed Minkowski space can be {\em ''naturally'' embedded} in a
five-dimensional (Riemannian) space [8].}. The generalized interval in $%
\widetilde{M_{4}}$ is\bigskip\ therefore given by ($x^{\mu
}=(x^{0},x^{1},x^{2},x^{3})=(ct,x,y,z)$, with $c$ being the usual\bigskip\
light speed in vacuum)\bigskip\
\begin{eqnarray}
ds^{2}(x^{5})
&=&b_{0}^{2}(x^{5})c^{2}dt^{2}-(b_{1}^{2}(x^{5})dx^{2}+b_{2}^{2}(x^{5})dy^{2}+b_{3}^{2}(x^{5})dz^{2})=
\nonumber \\
&&  \nonumber \\
&=&g_{\mu \nu ,DSR4}(x^{5})dx^{\mu }dx^{\nu }=dx\ast dx.
\end{eqnarray}
\bigskip

The last step in (2) defines the ($x^{5}$-dependent) scalar product $\ast $
in the deformed Minkowski\bigskip\ space $\widetilde{M_{4}}$ .\bigskip\ In
order to emphasize the dependence of DMS4 on the variable $x^{5}$, we shall
sometimes use the notation $\widetilde{M_{4}}(x^{5})$. It follows
immediately \bigskip that it can be regarded as a particular case of a
Riemann space with null curvature.\bigskip

From the general condition
\begin{equation}
g_{\mu \nu ,DSR4}(x^{5})g_{DSR4}^{\nu \rho }(x^{5})=\delta _{\mu }^{~~\rho }
\end{equation}
we get for the contravariant components of the metric tensor
\begin{eqnarray}
g_{DSR4}^{\mu \nu }(x^{5})
&=&diag(b_{0}^{-2}(x^{5}),-b_{1}^{-2}(x^{5}),-b_{2}^{-2}(x^{5}),-b_{3}^{-2}(x^{5}))=
\nonumber \\
&&  \nonumber \\
&&\stackrel{\text{{\footnotesize ESC off}}}{=}\delta ^{\mu \nu }\left(
b_{0}^{-2}(x^{5})\delta ^{\mu 0}-b_{1}^{-2}(x^{5})\delta ^{\mu
1}-b_{2}^{-2}(x^{5})\delta ^{\mu 2}-b_{3}^{-2}(x^{5})\delta ^{\mu 3}\right) .
\nonumber \\
&&
\end{eqnarray}
\bigskip

Let us stress that metric (1) is supposed to hold at a {\em local} (and not
global)\bigskip\ scale. We shall therefore refer to it as a ``{\em topical}%
'' deformed metric, because it\bigskip\ is supposed to be valid not
everywhere, but only in a suitable (local) \bigskip space-time region
(characteristic of both the system and the interaction considered).\bigskip

The two basic postulates of DSR4 (which generalize those of
standard\bigskip\ SR) are [3]:\bigskip

\ \ 1- {\em Space-time properties: } Space-time is homogeneous, but space is
not\bigskip\ necessarily isotropic; a reference frame in which space-time is
endowed with\bigskip\ such properties is called a ''{\em topical'' reference
frame} (TIRF). Two TIRF's are\bigskip\ in general moving uniformly with
respect to each other (i.e., as in SR, they\bigskip\ are connected by a
''inertiality'' relation, which defines an equivalence class\bigskip\ of $%
\infty ^{3}$ TIRF );\bigskip

2- {\em Generalized Principle of Relativity }(or {\em Principle of Metric
Invariance}):\bigskip\ All physical measurements within each TIRF must be
carried out via the\bigskip\ {\em same} metric.\bigskip

The metric (1) is just a possible realization of the above postulates.
We\bigskip\ refer the reader to Refs. [3]-[7] for the explicit expressions
of the \bigskip phenomenological energy-dependent metrics for the four
fundamental interactions\footnote{%
Since the metric coefficients $b_{\mu }^{2}(x^{5})$ are {\em dimensionless},
they actually do depend on the ratio $\frac{x^{5}}{x_{0}^{5}}$, where
\[
x_{0}^{5}\equiv \ell _{0}E_{0}
\]
is a {\em fundamental length}, proportional (by the {\em %
dimensionally-transposing} constant $\ell _{0}$) to the {\em threshold energy%
} $E_{0}$, characteristic of the interaction considered (see Refs. [3]-[8]).}%
.\bigskip\ \

\section{MAXIMAL\ KILLING\ GROUP\ OF\ GENERALIZED MINKOWSKI SPACES}

A $N$-dimensional {\em generalized Minkowski space }$\widetilde{M_{N}}%
(\left\{ x\right\} _{n.m.})$ is a Riemann space endowed with the global
metric structure [1]
\begin{equation}
ds^{2}=g_{\mu \nu }(\left\{ x\right\} _{n.m.})dx^{\mu }dx^{\nu }
\end{equation}
where the (in general non-diagonal) metric tensor $g_{\mu \nu }(\left\{
x\right\} _{n.m.})$ ($\mu ,\nu =1,2,...,N$) depends on a set $\left\{
x\right\} _{n.m.}$ of $N_{n.m.}$ non-metrical coordinates (i.e. different
from the $N$ coordinates related to the dimensions of the space considered).
We shall assume the (not necessarily hyperbolic) metric signature $(T,S)$ ($%
T $ timelike dimensions and $S=N-T$ spacelike dimensions). It follows that $%
\widetilde{M_{N}}(\left\{ x\right\} _{n.m.})$ is {\em flat}, because all the
components of the Riemann-Christoffel tensor vanish.

Of course, an example is just provided by the 4-d. deformed
Minkowski\bigskip\ space $\widetilde{M_{4}}(x^{5})$ discussed in the
previous Section.

The general form of the $N(N+1)/2$ Killing equations in a $N$-dimensional
generalized Minkowski space is given by [1] ($\mu $, $\nu =1,2,...,N$, and $%
{\bf x}$ denotes the usual contravariant coordinate $N$-vector):
\begin{equation}
\xi _{\mu }({\bf x})_{,\nu }+\xi _{\nu }({\bf x})_{,\mu }=0\equiv \frac{%
\partial \xi _{\mu }({\bf x})}{\partial x^{\nu }}+\frac{\partial \xi _{\nu }(%
{\bf x})}{\partial x^{\mu }}=0.
\end{equation}
Eqs. (6) are trivially satisfied by constant, covariant $N$-vectors $\xi
_{\mu }$ \ --- corresponding to the infinitesimal tranformation vectors of
the space-time, $N$-parameter translation (Lie) group $Tr.(S,T=N-S)_{GEN.}$
of the generalized Minkowskian space considered --- entering the general
expression of an infinitesimal translation [1]

\begin{gather}
tr.(S,T=N-S)_{GEN.}\ni \delta g:  \nonumber \\
\nonumber \\
x^{\mu }\rightarrow \left( x^{\prime }\right) ^{\mu }({\bf x},\left\{
x\right\} _{n.m.})=\left( x^{\mu }\right) ^{\prime }({\bf x},\left\{
x\right\} _{n.m.})=  \nonumber \\
\nonumber \\
=x^{\mu }+\delta x_{(g)}^{\mu }(\left\{ x\right\} _{n.m.})=x^{\mu }+\xi
_{(g)}^{\mu }(\left\{ x\right\} _{n.m.}),
\end{gather}
where $tr.(S,T=N-S)_{GEN.}$ is the (Lie) algebra of $N$-d. space-time
translations. Here, the contravariant $N$-vector $\delta x_{(g)}^{\mu
}(\left\{ x\right\} _{n.m.})=\xi _{(g)}^{\mu }(\left\{ x\right\} _{n.m.})$
is constant, i.e. independent of $x^{\mu }$.

Therefore, the maximal Killing group of $\widetilde{M_{N}}(\left\{ x\right\}
_{n.m.})$ is the {\em generalized Poincar\'{e}} (or {\em inhomogeneous
Lorentz}) {\em group} $P(S,T)_{GEN.}^{N(N+1)/2}$

\begin{equation}
P(T,S)_{GEN.}^{N(N+1)/2\text{{\footnotesize \ }}}=SO(T,S)_{GEN.}^{N(N-1)/2}%
\otimes _{s}Tr.(T,S)_{GEN.}^{N\text{{\small \ }}},
\end{equation}
i.e. the (semidirect \footnote{%
As already pointed out in Ref. [2] and as shall be explicitely derived (in
the hyperbolically-signed case $N=4,S=3,T=1$ of DSR4, without loss of
generality) in Subsect. 5.2, in general we have that
\[
\exists \text{ at least 1 }\left( \mu ,\nu ,\rho \right) \in \left\{
1,...,N\right\} ^{3}:[I_{GEN.}^{\mu \nu }(\left\{ x\right\}
_{n.m.}),\Upsilon _{GEN.}^{\rho }(\left\{ x\right\} _{n.m.})]\neq 0,\text{ }%
\forall \left\{ x\right\} _{n.m.},
\]
where $I_{GEN.}^{\mu \nu }(\left\{ x\right\} _{n.m.})$ are generalized
rotation infinitesimal generators (i.e. generators of $%
SO(T,S)_{GEN.}^{N(N-1)/2}$) and $\Upsilon _{GEN.}^{\rho }(\left\{ x\right\}
_{n.m.})$ are generalized translation infinitesimal generators (i.e.
generators of $Tr.(T,S)_{GEN.}^{N\text{{\small \ }}}$). Therefore, the
correct group product to be considered is the semidirect one (see e.g. Ref.
[10]).}) product of the Lie group of $N$-dimensional generalized space-time
rotations (or $N$-d. generalized, homogeneous Lorentz group $%
SO(T,S)_{GEN.}^{N(N-1)/2\text{{\footnotesize \ }}}$) with $N(N-1)/2$
parameters, and of the Lie group of generalized $N$-dimensional space-time
translations $Tr.(T,S)_{GEN.}^{N}$ with $N$ parameters (see Ref. [1]).

\section{TRANSLATIONS\ IN\ 4-d. GENERALIZED MINKOWSKI SPACES}

In the case $N=4$, one can write the components of the covariant Killing
4-vector of a generic 4-d. generalized Minkowski space $\widetilde{M_{4}}%
(\left\{ x\right\} _{n.m.})$ as [1] (by omitting, for simplicity's sake, the
dependence on the group element $g\in Tr.(S,T=4-S)_{GEN.}\subset
P(S,T=4-S)_{GEN.}$)
\begin{equation}
\left\{
\begin{array}{c}
\xi _{0}(\left\{ x\right\} _{m.})=-\zeta ^{1}x^{1}-\zeta ^{2}x^{2}-\zeta
^{3}x^{3}+T^{0}\medskip , \\
\\
\xi _{1}(\left\{ x\right\} _{m.})=\zeta ^{1}x^{0}+\theta ^{2}x^{3}-\theta
^{3}x^{2}-T^{1}\medskip , \\
\\
\xi _{2}(\left\{ x\right\} _{m.})=\zeta ^{2}x^{0}-\theta ^{1}x^{3}+\theta
^{3}x^{1}-T^{2}\medskip , \\
\\
\xi _{3}(\left\{ x\right\} _{m.})=\zeta ^{3}x^{0}+\theta ^{1}x^{2}-\theta
^{2}x^{1}-T^{3}\medskip .
\end{array}
\right.
\end{equation}
Thus, independently of the explicit form of the metric tensor, {\em all 4-d.
generalized Minkowski spaces admit the same covariant Killing vector}. In
particular, with the hyperbolic signature $(+,-,-,-)$ (namely $S=3$, $T=1$)
of SR and DSR4, it can be shown ([1],[2]) that ${\bf \zeta }=(\zeta
^{1},\zeta ^{2},\zeta ^{3})$ is the (Euclidean) 3-vector of the
dimensionless parameters (i.e. ''generalized rapidities'') of a generalized
3-d. ''boosts'' and ${\bf \theta }=(\theta ^{1},\theta ^{2},\theta ^{3})$ is
the (Euclidean) 3-vector of the dimensionless parameters (i.e. generalized
angles) of a generalized 3-d. true rotations, whereas
\begin{equation}
T_{\mu }=(T^{0},-T^{1},-T^{2},-T^{3})
\end{equation}
is the covariant 4-vector of the length-dimensioned parameters of a
generalized 4-d. translation \footnote{%
Let us notice an important fact, peculiar to the translation component $%
Tr.(3,1)_{GEN.}$ of the 4-d. $(3,1)$ generalized Poincar\`{e} group $%
P(3,1)_{GEN.}$.
\par
The elements of the 4-d. $(3,1)$ generalized rotation group $SO(3,1)_{GEN.}$
correspond, both at infinitesimal [1] and finite [2] level, to coordinate
transformations homogeneous in their arguments, i.e. in the
''length-dimensioned'' coordinate basis $\left\{ x^{\mu }\right\} _{\mu
=0,1,2,3}$. It is then clear that the generalized parametric (Euclidean)
3-vectors $\underline{\theta }(g)$\ and $\underline{\zeta }(g)$\ must be
dimensionless. In the cases $S=3,T=1$ of SR (corresponding to $M_{4}$) and
of DSR4 ([1],[2]) (corresponding to $\widetilde{M_{4}}\left( x^{5}\right) $)
$\underline{\theta }(g)$\ and $\underline{\zeta }(g)$ have been identified
with the generalized true rotation angle and generalized ''boost
rapidities'' 3-vectors, respectively. By using the dimension-transposing
constant velocity $c$, it has been possible to introduce a
''velocity-dimensioned'' ''boost'' parametric 3-vector $\underline{v}(g)$,
that is a contravariant 3-vector in the 3-d. physical space embedded in the
4-d. Minkowski space being considered ($E_{3}\subset M_{4}$ in SR, and $%
\widetilde{E_{3}}\left( x^{5}\right) \subset \widetilde{M_{4}}\left(
x^{5}\right) $ in DSR4, respectively).
\par
Analogously, because of the fact that the elements of the 4-d. $(3,1)$
generalized translation group $Tr.(3,1)_{GEN.}$ correspond, both at
infinitesimal and finite level, to purely inhomogeneous coordinate
transformations, it is clear that\ the generalized translation parametric
4-vector $T^{\mu }(g)$\ must be ''length-dimensioned'' and have a {\em %
''context-dependent''} geometric nature. For example it is a ''standard''
contravariant 4-vector $T_{SR}^{\mu }(g)$ of $M_{4}$ in SR, and a
''deformed'' contravariant 4-vector $T_{DSR4}^{\mu }(g)$ of $\widetilde{M_{4}%
}\left( x^{5}\right) $ in DSR4.
\par
That is why 3-d. Euclidean scalar products $\underline{\theta }(g)\cdot
\underline{S_{DSR4}}(x^{5})$\ and $\underline{\zeta }(g)\cdot \underline{%
K_{DSR4}}(x^{5})$,\ and 4-d. ''deformed'' scalar product (ESC on)
\[
T_{\mu ,\left( DSR4\right) }(g)\Upsilon _{DSR4}^{\mu }(x^{5})=\Upsilon
_{DSR4}^{\mu }(x^{5})g_{\mu \nu ,DSR4}(x^{5})T_{DSR4}^{\nu }(g,x^{5})
\]
do appear in Eq. (53), which expresses the general form of the $5\times 5$\
matrix corresponding to a finite transformation of the 4-d. ''deformed''
(inhomogeneous Lorentz) Poincar\`{e} group $P(3,1)_{DEF.}$.
\par
Notice that the notation ''$(DSR4)$'' in $T_{\mu ,\left( DSR4\right) }$ has
been used to mean that actually, as expressed by Eq. (10), $T_{\mu }(g)$\ is
independent of the (4-d.) metric context being considered.}.

The inhomogeneity of the (infinitesimal) translation transformation (7)
obviously implies that it {\em cannot} be represented by a $4\times 4$
matrix (at the infinitesimal, and then at the finite, level), i.e. no 4-d.
representation of the infinitesimal generators of $Tr.(3,1)_{GEN.}$ exists.
However, it is possible to get a matrix representation of the infinitesimal
generators of the generalized translation group $Tr.(S,T=4-S)_{GEN.}\subset
P(S,T=4-S)_{GEN.}$ by introducing a fifth auxiliary coordinate [9] $y=1$,
devoid of any physical or metric meaning. This fictitious extra coordinate
is introduced to the only aim of parametrizing the non-homogeneous part of
the coordinate transformations of 4-d. $(3,1)$ generalized Poincar\'{e}
group $P(3,1)_{GEN.}$\footnote{%
Let us stress that the coordinate $y=1$\ has a merely parametrizing meaning,
that is it has to span the {\em ''transformative degree of freedom''}
associated to the inhomogeneous component of the (maximal) Killing group $%
P(3,1)_{GEN.}$\ of the 4-d. $(3,1)$ generalized Minkowski space $\widetilde{%
M_{4}}\left( \left\{ x\right\} _{n.m.}\right) $ being considered. In other
words, $y=1$\ has to express the translation component of the Poincar\`{e}
generalized coordinate transformations of $\widetilde{M_{4}}\left( \left\{
x\right\} _{n.m.}\right) $.
\par
The coordinate $y$ has also a null total differential, because it is
constant:
\[
y=1\Rightarrow dy=0,
\]
whence it has not physical nor metric meaning.
\par
Moreover, the trivial process of {\em ''dimensional embedding''} (4-d.$%
\rightarrow $5-d.) of the (matrix) representation of infinitesimal
generators of $SO(3,1)_{GEN.}$\ group (as expressed by Eq. (28)), does {\em %
not} change the infinitesimal-algebraic structure in any way; this is
because of the fact that matrix rows and columns corresponding to the
''auxiliary coordinate'' $y$ do {\em not} ''mix'' with\ the homogeneous
components of the coordinate transformations being considered.
\par
\ More generally, the introduction of $y$\ is necessary to give an explicit $%
\left( N+1\right) $-d. (matrix) representation of the infinitesimal
generators of the generalized translation group $Tr.(S,T=N-S)_{GEN.}$, and
then to calculate the (representation-independent) $N(S,T)$ generalized
''mixed'' Poincar\`{e} algebra, i.e. the commutator-exploited algebraic
structure between the infinitesimal generators of $SO(S,T=N-S)_{GEN.}$ and
the infinitesimal generators of $Tr.(S,T=N-S)_{GEN.}$\ (for the case $%
N=4,S=3,T=1$ of DSR4, see Subsect. 5.2).}.

Then, following the notation of page 150 of Ref. [9]\footnote{%
The only difference with Ref. [9] (treating the SR case) is an {\em overall}
minus sign. This is fully justifiable assuming that the parametric
contravariant 4-vector $\varepsilon ^{\mu }$, used in Eq. (6-5.35) of page
150 in Ref. [1], is the {\em opposite} of $T_{SSR4}^{\mu }$; that is, by
omitting, for simplicity's sake, the dependence on $g\in
Tr(3,1)_{STD.}\subset P(3,1)_{STD.}$:
\[
\varepsilon ^{\mu }\equiv -T_{SSR4}^{\mu }=\left(
-T^{0},-T^{1},-T^{2},-T^{3}\right) .
\]
\
\par
{}}, one can consider\footnote{%
This choice could now seem a bit arbitrary, but it will prove to be
justified and self-consistent from the following results, obtained, without
loss of generality, in the case of DSR4 (see also Footnote 10).} the
following 5-d. (matrix) representation of the infinitesimal generators%
\footnote{%
In the following, the upper-case Latin indices have range $\left\{
0,1,2,3,6\right\} $,{\footnotesize \ }where the index $6$ labels the
auxiliary coordinate:
\[
x^{6}\equiv x_{6}=y=1
\]
\par
Moreover, independently of the contravariant or covariant nature of
infinitesimal generators, contravariant and covariant indices in their
(matrix) representations conventionally stand for row and column indices,
respectively.} $\left\{ \left( \Upsilon _{\mu }\right) _{~~B}^{A}\right\}
_{\mu =0,1,2,3}$ of the group $Tr.(3,1)_{GEN.}$:
\begin{equation}
\begin{array}{cc}
\Upsilon _{0}\equiv \left(
\begin{array}{ccccc}
0 & 0 & 0 & 0 & 1 \\
0 & 0 & 0 & 0 & 0 \\
0 & 0 & 0 & 0 & 0 \\
0 & 0 & 0 & 0 & 0 \\
0 & 0 & 0 & 0 & 0
\end{array}
\right) ; & \Upsilon _{1}\equiv \left(
\begin{array}{ccccc}
0 & 0 & 0 & 0 & 0 \\
0 & 0 & 0 & 0 & 1 \\
0 & 0 & 0 & 0 & 0 \\
0 & 0 & 0 & 0 & 0 \\
0 & 0 & 0 & 0 & 0
\end{array}
\right) ;
\end{array}
\end{equation}
\begin{equation}
\begin{array}{cc}
\Upsilon _{2}\equiv \left(
\begin{array}{ccccc}
0 & 0 & 0 & 0 & 0 \\
0 & 0 & 0 & 0 & 0 \\
0 & 0 & 0 & 0 & 1 \\
0 & 0 & 0 & 0 & 0 \\
0 & 0 & 0 & 0 & 0
\end{array}
\right) ; & \Upsilon _{3}\equiv \left(
\begin{array}{ccccc}
0 & 0 & 0 & 0 & 0 \\
0 & 0 & 0 & 0 & 0 \\
0 & 0 & 0 & 0 & 0 \\
0 & 0 & 0 & 0 & 1 \\
0 & 0 & 0 & 0 & 0
\end{array}
\right) .
\end{array}
\end{equation}
Namely, the only non-zero components of the above 5-d. representative
matrices are
\begin{equation}
\left( \Upsilon _{0}\right) _{~~6}^{0}=\left( \Upsilon _{1}\right)
_{~~6}^{1}=\left( \Upsilon _{2}\right) _{~~6}^{2}=\left( \Upsilon
_{3}\right) _{~~6}^{3}=1,
\end{equation}
or, equivalently:
\begin{equation}
\left( \Upsilon _{\mu }\right) _{~~B}^{A}=\delta _{\mu }^{~~A}\delta _{6B}.
\end{equation}

From Eqs. (11) and (12) one easily find the following properties of the
above considered 5-d. representation of the {\em covariant} infinitesimal
generators of $Tr.(3,1)_{GEN.}$ (here and in the following, $0_{5-d.}$ and $%
1_{5-d.}$ denote the $5\times 5$ zero and unity matrix, respectively):
\begin{gather}
\left( \Upsilon _{\mu }\right) ^{n}=0_{5-d.},\forall n\geq 2\Rightarrow \exp
(\Upsilon _{\mu })=1_{5-d.}+\Upsilon _{\mu }\medskip ; \\
\nonumber \\
\lbrack \Upsilon _{\mu },\Upsilon _{\nu }]=0_{5-d.},\text{ \ }\forall \left(
\mu ,\nu \right) \in \left\{ 0,1,2,3\right\} ^{2}\medskip ; \\
\nonumber \\
\Upsilon _{\mu }\neq \Upsilon _{\mu }\left( \left\{ x\right\} _{m.},\left\{
x\right\} _{n.m.}\right) \medskip .
\end{gather}

In the following, we shall see that, in the DSR4 case, the properties (15)
and (16) still hold for 5-d. matrix representation of the {\em contravariant}
infinitesimal deformed translation generators. Moreover, as it is clear from
Eqs. (11)-(14), the considered 5-d. representation of the {\em covariant}
infintesimal generators of $Tr.(3,1)_{GEN.}$ are independent of the metric
tensor (namely, they are the same irrespective of the 4-d. generalized
Minkowski space $\widetilde{M_{4}}(\left\{ x\right\} _{n.m.})$ considered).
On the contrary, the {\em contravariant} generators {\em do depend} on the
generalized metric, since\footnote{%
The assumed explicit 5-d. representations (11) and (12) are justifiable with
the following reasoning. Eq. (9) (see [1]) expresses the independence of $%
T_{\mu }(g)$\ on the (geo)metric context being considered; instead, its {\em %
contravariant} form will in general be ''context-dependent'', of course (see
Footnote 5):
\[
T^{\mu }(g,\left\{ x\right\} _{n.m.})\stackrel{\text{ESC on}}{\equiv }g^{\mu
\nu }(\left\{ x\right\} _{n.m.})T_{\nu }(g).
\]
\par
Whence, because of the fact that in general a 4-d. ''context-dependent''
scalar product
\[
T_{\mu }(g)\Upsilon ^{\mu }(\left\{ x\right\} _{n.m.})=T_{\mu }(g)g^{\mu \nu
}(\left\{ x\right\} _{n.m.})\Upsilon _{\nu }=T^{\mu }(g,\left\{ x\right\}
_{n.m.})\Upsilon _{\mu }
\]
appear in the explicit form of 5-d. matrix of infinitesimal generalized
translation (e.g., in DSR4 case, $T_{T_{DSR4}^{\mu }(g,x^{5}),DSR4}(x^{5})$\
of Eq. (48)), it is then clear that the set of the {\em covariant}
infinitesimal generalized translation generators has to be {\em %
''context-independent}'', i.e. has to be the same irrespective of the 4-d.
generalized Minkowski space $\widetilde{M_{4}}(\left\{ x\right\} _{n.m.})$
considered.}
\begin{equation}
\Upsilon ^{\mu }\stackrel{\text{ESC on}}{=}g^{\mu \rho }(\left\{ x\right\}
_{n.m.})\Upsilon _{\rho }=\Upsilon ^{\mu }(\left\{ x\right\} _{n.m.}).
\end{equation}

\section{\ THE\ GROUP $Tr.(3,1)_{DEF.}$ OF 4-d. DEFORMED TRANSLATIONS IN $%
\widetilde{M_{4}}(x^{5})$}

\subsection{The 5-d. representation of infinitesimal contravariant generators%
}

Let us consider the case of $Tr.(3,1)_{DEF.}$, i.e. the deformed space-time
translation group of the 4-d. deformed Minkowski space $\widetilde{M_{4}}%
(x^{5})$ of DSR4, whose metric tensor is given by Eq.(3). Then, on account
of Eqs. (11)-(12) and (18), the considered 5-d. matrix representation of the
infinitesimal {\em contravariant} deformed translation generators reads
\begin{equation}
\Upsilon _{DSR4}^{0}(x^{5})\equiv \left(
\begin{array}{ccccc}
0 & 0 & 0 & 0 & b_{0}^{-2}(x^{5}) \\
0 & 0 & 0 & 0 & 0 \\
0 & 0 & 0 & 0 & 0 \\
0 & 0 & 0 & 0 & 0 \\
0 & 0 & 0 & 0 & 0
\end{array}
\right) ;
\end{equation}
\begin{equation}
\Upsilon _{DSR4}^{1}(x^{5})\equiv \left(
\begin{array}{ccccc}
0 & 0 & 0 & 0 & 0 \\
0 & 0 & 0 & 0 & -b_{1}^{-2}(x^{5}) \\
0 & 0 & 0 & 0 & 0 \\
0 & 0 & 0 & 0 & 0 \\
0 & 0 & 0 & 0 & 0
\end{array}
\right) ;
\end{equation}
\begin{equation}
\Upsilon _{DSR4}^{2}(x^{5})\equiv \left(
\begin{array}{ccccc}
0 & 0 & 0 & 0 & 0 \\
0 & 0 & 0 & 0 & 0 \\
0 & 0 & 0 & 0 & -b_{2}^{-2}(x^{5}) \\
0 & 0 & 0 & 0 & 0 \\
0 & 0 & 0 & 0 & 0
\end{array}
\right) ;
\end{equation}
\begin{equation}
\Upsilon _{DSR4}^{3}(x^{5})\equiv \left(
\begin{array}{ccccc}
0 & 0 & 0 & 0 & 0 \\
0 & 0 & 0 & 0 & 0 \\
0 & 0 & 0 & 0 & 0 \\
0 & 0 & 0 & 0 & -b_{3}^{-2}(x^{5}) \\
0 & 0 & 0 & 0 & 0
\end{array}
\right) .
\end{equation}
The only non-zero components are therefore given by
\begin{eqnarray}
\left( \Upsilon _{DSR4}^{0}(x^{5})\right) _{~~6}^{0} &=&-\frac{%
b_{0}^{-2}(x^{5})}{b_{1}^{-2}(x^{5})}\left( \Upsilon
_{DSR4}^{1}(x^{5})\right) _{~~6}^{1}=  \nonumber \\
&&  \nonumber \\
&&  \nonumber \\
&=&-\frac{b_{0}^{-2}(x^{5})}{b_{2}^{-2}(x^{5})}\left( \Upsilon
_{DSR4}^{2}(x^{5})\right) _{~~6}^{2}=-\frac{b_{0}^{-2}(x^{5})}{%
b_{3}^{-2}(x^{5})}\left( \Upsilon _{DSR4}^{3}(x^{5})\right)
_{~~6}^{3}=b_{0}^{-2}(x^{5}),\medskip  \nonumber \\
&&
\end{eqnarray}
or equivalently (ESC off) :
\begin{equation}
\left( \Upsilon _{DSR4}^{\mu }(x^{5})\right) _{~~B}^{A}=\left(
b_{0}^{-2}(x^{5})\delta ^{\mu 0}-b_{1}^{-2}(x^{5})\delta ^{\mu
1}-b_{2}^{-2}(x^{5})\delta ^{\mu 2}-b_{3}^{-2}(x^{5})\delta ^{\mu 3}\right)
\delta _{\mu }^{~~A}\delta _{6B}.
\end{equation}

From Eqs. (19)-(22) one immediately gets the following ({\em %
representation-independent}) properties of the {\em contravariant} deformed
translation infinitesimal generators in $\widetilde{M_{4}}(x^{5})$:
\begin{eqnarray}
\left( \Upsilon _{DSR4}^{\mu }(x^{5})\right) ^{n} &=&0_{5-d.},\forall n\geq
2\Rightarrow \exp (\Upsilon _{DSR4}^{\mu }(x^{5}))=1_{5-d.}+\Upsilon
_{DSR4}^{\mu }(x^{5});  \nonumber \\
&&
\end{eqnarray}
\begin{eqnarray}
\lbrack \Upsilon _{DSR4}^{\mu }(x^{5}),\Upsilon _{DSR4}^{\nu }(x^{5})]
&=&0_{5-d.},\forall \mu ,\nu \in \left\{ 0,1,2,3\right\} ; \\
&&  \nonumber
\end{eqnarray}
\begin{eqnarray}
\Upsilon _{DSR4}^{\mu } &=&\Upsilon _{DSR4}^{\mu }(x^{5}),\Upsilon
_{DSR4}^{\mu }\neq \Upsilon _{DSR4}^{\mu }(\left\{ x\right\} _{m.}). \\
&&  \nonumber
\end{eqnarray}
It follows from Eq. (26) that ($tr.(3,1)_{DEF.}$) $Tr.(3,1)_{DEF.}$ is a
proper (subalgebra) Abelian subgroup of the 4-d. deformed Poincar\'{e}
(algebra) group ($su(2)_{DEF.}\times su(2)_{DEF.}\times _{s}tr.(3,1)_{DEF.}$%
) $P(3,1)_{DEF.}$, whose infinitesimal ({\em contravariant}) generators (by
Eq. (27)) are {\em independent} of the metric variables of $\widetilde{M_{4}}%
(x^{5})$, \ but do ({\em parametrically}) depend on the {\em non-metric}
variable $x^{5}$.

\subsection{The ''mixed'' deformed Poincar\`{e} algebra and deformed
Poincar\`{e} algebra $\left( su(2)_{DEF.}\times su(2)_{DEF.}\right) \times
_{s}tr.(3,1)_{DEF.}$}

It is now possible to find the ''mixed'' algebraic structure of the 4-d.
deformed Poincar\'{e} group $P(3,1)_{DEF.}$ of DSR4; this can be exploited
by evaluating the commutators (which, as in the case of the 4-d. deformed
Lorentz algebra [1] $su(2)_{DEF.}\times su(2)_{DEF.}$, will be {\em %
representation-independent}) among the infinitesimal generators of $%
Tr.(3,1)_{DEF.}$ and the infinitesimal generators of the deformed
homogeneous Lorentz group $SO(3,1)_{DEF.}$. To this aim, one has to
represent the infinitesimal generators of $SO(3,1)_{DEF.}$ as $5\times 5$
matrices in the auxiliary fictitious 5-d. space with $y=1$ as extra
dimension. It is easy to see that this amounts to the following trivial
replacement:
\begin{equation}
I_{DSR4}^{\alpha \beta }(x^{5})\rightarrow \left(
\begin{array}{cc}
I_{DSR4}^{\alpha \beta }(x^{5}) & 0 \\
0 & 0
\end{array}
\right) \text{ \ }\forall \left( \alpha ,\beta \right) \in \left\{
0,1,2,3\right\} ^{2},
\end{equation}
where $I_{DSR4}^{\alpha \beta }(x^{5})$ are the infinitesimal generators of
the 4-d. deformed homogeneous Lorentz group $SO(3,1)_{DEF.}$ of DSR4 in the
4-d. matrix representation derived in Ref. [1]. We have therefore:
\begin{equation}
\begin{array}{cc}
I_{DSR4}^{10}(x^{5})= & \left(
\begin{array}{ccccc}
0 & -b_{0}^{-2}(x^{5}) & 0 & 0 & 0 \\
-b_{1}^{-2}(x^{5}) & 0 & 0 & 0 & 0 \\
0 & 0 & 0 & 0 & 0 \\
0 & 0 & 0 & 0 & 0 \\
0 & 0 & 0 & 0 & 0
\end{array}
\right) ;
\end{array}
\end{equation}
\begin{equation}
\begin{array}{cc}
I_{DSR4}^{20}(x^{5})= & \left(
\begin{array}{ccccc}
0 & 0 & -b_{0}^{-2}(x^{5}) & 0 & 0 \\
0 & 0 & 0 & 0 & 0 \\
-b_{2}^{-2}(x^{5}) & 0 & 0 & 0 & 0 \\
0 & 0 & 0 & 0 & 0 \\
0 & 0 & 0 & 0 & 0
\end{array}
\right) ;
\end{array}
\end{equation}
\begin{equation}
\begin{array}{cc}
I_{DSR4}^{30}(x^{5})= & \left(
\begin{array}{ccccc}
0 & 0 & 0 & -b_{0}^{-2}(x^{5}) & 0 \\
0 & 0 & 0 & 0 & 0 \\
0 & 0 & 0 & 0 & 0 \\
-b_{3}^{-2}(x^{5}) & 0 & 0 & 0 & 0 \\
0 & 0 & 0 & 0 & 0
\end{array}
\right) ;
\end{array}
\end{equation}
\begin{equation}
\begin{array}{cc}
I_{DSR4}^{12}(x^{5})= & \left(
\begin{array}{ccccc}
0 & 0 & 0 & 0 & 0 \\
0 & 0 & -b_{1}^{-2}(x^{5}) & 0 & 0 \\
0 & b_{2}^{-2}(x^{5}) & 0 & 0 & 0 \\
0 & 0 & 0 & 0 & 0 \\
0 & 0 & 0 & 0 & 0
\end{array}
\right) ;
\end{array}
\end{equation}
\begin{equation}
\begin{array}{cc}
I_{DSR4}^{23}(x^{5})= & \left(
\begin{array}{ccccc}
0 & 0 & 0 & 0 & 0 \\
0 & 0 & 0 & 0 & 0 \\
0 & 0 & 0 & -b_{2}^{-2}(x^{5}) & 0 \\
0 & 0 & b_{3}^{-2}(x^{5}) & 0 & 0 \\
0 & 0 & 0 & 0 & 0
\end{array}
\right) ;
\end{array}
\end{equation}
\begin{equation}
\begin{array}{cc}
I_{DSR4}^{31}(x^{5})= & \left(
\begin{array}{ccccc}
0 & 0 & 0 & 0 & 0 \\
0 & 0 & 0 & b_{1}^{-2}(x^{5}) & 0 \\
0 & 0 & 0 & 0 & 0 \\
0 & -b_{3}^{-2}(x^{5}) & 0 & 0 & 0 \\
0 & 0 & 0 & 0 & 0
\end{array}
\right) .
\end{array}
\end{equation}

From Eqs. (19)-(22) and (29)-(34) one gets the following form for the 4-d.
''mixed'' deformed Poincar\`{e} algebra ($\forall \left( i,j,k\right) \in
\left\{ 1,2,3\right\} ^{3}$) :
\begin{eqnarray}
&&\left\{
\begin{array}{l}
\lbrack I_{DSR4}^{i0}(x^{5}),\Upsilon
_{DSR4}^{0}(x^{5})]=b_{0}^{-2}(x^{5})\Upsilon _{DSR4}^{i}(x^{5})\medskip ;
\\
\\
\lbrack I_{DSR4}^{i0}(x^{5}),\Upsilon _{DSR4}^{j}(x^{5})]\stackrel{\text{ESC
off on }i}{=}\delta ^{ij}(x^{5})b_{i}^{-2}(x^{5})\Upsilon
_{DSR4}^{0}(x^{5})\medskip ; \\
\\
\lbrack I_{DSR4}^{ij}(x^{5}),\Upsilon _{DSR4}^{0}(x^{5})]=0\medskip ; \\
\\
\lbrack I_{DSR4}^{ij}(x^{5}),\Upsilon _{DSR4}^{k}(x^{5})]\stackrel{\text{ESC
off on }i\text{ and }j}{=}\delta ^{ik}b_{i}^{-2}(x^{5})\Upsilon
_{DSR4}^{j}(x^{5})-\delta ^{jk}b_{j}^{-2}(x^{5})\Upsilon
_{DSR4}^{i}(x^{5})\medskip ,
\end{array}
\right.  \nonumber \\
&&
\end{eqnarray}
or in compact form ($\forall \left( \mu ,\nu ,\rho \right) \in \left\{
0,1,2,3\right\} ^{3}$):
\begin{gather}
\lbrack I_{DSR4}^{\mu \nu }(x^{5}),\Upsilon _{DSR4}^{\rho }(x^{5})]=
\nonumber \\
\nonumber \\
=g_{DSR4}^{\nu \rho }(x^{5})\Upsilon _{DSR4}^{\mu }(x^{5})-g_{DSR4}^{\mu
\rho }(x^{5})\Upsilon _{DSR4}^{\nu }(x^{5})=\medskip  \nonumber \\
\nonumber \\
\stackrel{\text{{\footnotesize ESC off}}}{=}\delta ^{\nu \rho }\left(
b_{0}^{-2}(x^{5})\delta ^{\nu 0}-b_{1}^{-2}(x^{5})\delta ^{\nu
1}-b_{2}^{-2}(x^{5})\delta ^{\nu 2}-b_{3}^{-2}(x^{5})\delta ^{\nu 3}\right)
\Upsilon _{DSR4}^{\mu }(x^{5})+\medskip  \nonumber \\
\nonumber \\
-\delta ^{\mu \rho }\left( b_{0}^{-2}(x^{5})\delta ^{\mu
0}-b_{1}^{-2}(x^{5})\delta ^{\mu 1}-b_{2}^{-2}(x^{5})\delta ^{\mu
2}-b_{3}^{-2}(x^{5})\delta ^{\mu 3}\right) \Upsilon _{DSR4}^{\nu
}(x^{5})\medskip .
\end{gather}
Whence, in general
\begin{equation}
\exists \text{ at least 1 }\left( \mu ,\nu ,\rho \right) \in \left\{
0,1,2,3\right\} ^{3}:[I_{DSR4}^{\mu \nu }(x^{5}),\Upsilon _{DSR4}^{\rho
}(x^{5})]\neq 0,\text{ }\forall x^{5}\in R_{0}^{+}.
\end{equation}
Therefore, although ($su(2)_{DEF.}\times su(2)_{DEF.}$) $SO(3,1)_{DEF.}$ and
($tr.(3,1)_{DEF.}$) $Tr.(3,1)_{DEF.}$ are proper (subalgebras) subgroups -
non-Abelian and Abelian, respectively - of the 4-d. deformed Poincar\'{e}
(algebra) group [1], they determine it only by their semidirect product (see
e.g. Ref. [10]).

Let us change the basis of infinitesimal generators of $SO(3,1)_{DEF.}$ to
the ''self-representative'' one by defining the following deformed
space-time infinitesimal generator Euclidean 3-vectors ([1],[2]) ($\forall
i=1,2,3$):
\begin{gather}
S_{DSR4}^{i}(x^{5})\stackrel{\text{ESC on}}{\equiv }\frac{1}{2}\epsilon
_{~jk}^{i}I_{DSR4}^{jk}(x^{5}),\medskip \\
\nonumber \\
K_{DSR4}^{i}(x^{5})\equiv I_{DSR4}^{0i}(x^{5})\medskip ,
\end{gather}
where $\epsilon _{ijk}$\ is the (Euclidean) Levi-Civita 3-tensor with the
convention $\epsilon _{123}\equiv 1$. In this basis, the ''mixed'' part of
the 4-d. deformed Poincar\'{e} algebra can be written as (ESC off) :
\[
\lbrack K_{DSR4}^{i}(x^{5}),\Upsilon
_{DSR4}^{0}(x^{5})]=-b_{0}^{-2}(x^{5})\Upsilon _{DSR4}^{i}(x^{5})\bigskip ;
\]
\[
\lbrack K_{DSR4}^{i}(x^{5}),\Upsilon _{DSR4}^{j}(x^{5})]\stackrel{ESCoff%
\text{ }on\text{ }i}{=}-\delta ^{ij}b_{i}^{-2}(x^{5})\Upsilon
_{DSR4}^{0}(x^{5})\bigskip ;
\]
\begin{eqnarray*}
\lbrack S_{DSR4}^{i}(x^{5}),\Upsilon _{DSR4}^{0}(x^{5})] &=&[\frac{1}{2}%
\epsilon _{~jk}^{i}I_{DSR4}^{jk}(x^{5}),\Upsilon _{DSR4}^{0}](x^{5})=\medskip
\\
&& \\
&& \\
&=&\frac{1}{2}\epsilon _{~jk}^{i}[I_{DSR4}^{jk}(x^{5}),\Upsilon
_{DSR4}^{0}(x^{5})]=0\bigskip ;
\end{eqnarray*}
\begin{eqnarray}
\lbrack S_{DSR4}^{i}(x^{5}),\Upsilon _{DSR4}^{k}(x^{5})] &=&[\frac{1}{2}%
\epsilon _{~jl}^{i}I_{DSR4}^{jl}(x^{5}),\Upsilon _{DSR4}^{k}(x^{5})]=
\nonumber \\
&&  \nonumber \\
&&  \nonumber \\
&=&\frac{1}{2}\epsilon _{~jl}^{i}[I_{DSR4}^{jl}(x^{5}),\Upsilon
_{DSR4}^{k}(x^{5})]=\medskip  \nonumber \\
&&  \nonumber \\
&&  \nonumber \\
&=&\frac{1}{2}\epsilon _{~jl}^{i}\left( \delta
^{jk}b_{j}^{-2}(x^{5})\Upsilon _{DSR4}^{l}(x^{5})-\delta
^{lk}b_{l}^{-2}(x^{5})\Upsilon _{DSR4}^{j}(x^{5})\right) \stackrel{\text{ESC
off on }k}{=}  \nonumber \\
&&  \nonumber \\
&&  \nonumber \\
&=&\frac{1}{2}\left( \epsilon _{~~~l}^{ik}b_{k}^{-2}(x^{5})\Upsilon
_{DSR4}^{l}(x^{5})-\epsilon _{~~j}^{i~~~k}b_{k}^{-2}(x^{5})\Upsilon
_{DSR4}^{j}(x^{5})\right) =  \nonumber \\
&&  \nonumber \\
&&  \nonumber \\
&&\stackrel{\text{ESC off on }k}{=}\epsilon _{ikl}b_{k}^{-2}(x^{5})\Upsilon
_{DSR4}^{l}(x^{5})\medskip .  \nonumber \\
&&
\end{eqnarray}

On account of the results obtained in Ref. [1] for the 4-d. deformed
(homogeneous) Lorentz algebra $su(2)_{DEF.}\times su(2)_{DEF.}$, we can
write the whole Lie algebra $\left( su(2)_{DEF.}\times su(2)_{DEF.}\right)
\times _{s}tr.(3,1)_{DEF.}$ of the 4-d. deformed Poincar\'{e} group $%
P(3,1)_{DEF.}$ (i.e. the algebraic-infinitesimal structure of the maximal
Killing group of $\widetilde{M_{4}}(x^{5})$) as
\begin{gather}
\begin{array}{c}
\text{4-d. deformed space-time\medskip } \\
\text{rotation algebra }su(2)_{DEF.}\times su(2)_{DEF.}\text{\medskip\ :}
\end{array}
\nonumber \\
\nonumber \\
\nonumber \\
\begin{array}{l}
\lbrack I_{DSR4}^{\alpha \beta }(x^{5}),I_{DSR4}^{\rho \sigma
}(x^{5})]=\medskip \\
\\
\\
=g_{DSR4}^{\alpha \sigma }(x^{5})I_{DSR4}^{\beta \rho
}(x^{5})+g_{DSR4}^{\beta \rho }(x^{5})I_{DSR4}^{\alpha \sigma }(x^{5})+ \\
\medskip \\
-g_{DSR4}^{\alpha \rho }(x^{5})I_{DSR4}^{\beta \sigma
}(x^{5})-g_{DSR4}^{\beta \sigma }(x^{5})I_{DSR4}^{\alpha \rho }(x^{5})= \\
\medskip \\
\\
\stackrel{\text{{\footnotesize ESC off}}}{=}\delta ^{\alpha \sigma
}(b_{0}^{-2}(x^{5})\delta ^{\alpha 0}-b_{1}^{-2}(x^{5})\delta ^{\alpha
1}+\medskip \\
\\
-b_{2}^{-2}(x^{5})\delta ^{\alpha 2}-b_{3}^{-2}(x^{5})\delta ^{\alpha
3})I_{DSR4}^{\beta \rho }(x^{5})+\medskip \\
\\
+\delta ^{\beta \rho }(\delta ^{\beta 0}b_{0}^{-2}(x^{5})-\delta ^{\beta
1}b_{1}^{-2}(x^{5})+\medskip \\
\\
-\delta ^{\beta 2}b_{2}^{-2}(x^{5})-\delta ^{\beta
3}b_{3}^{-2}(x^{5}))I_{DSR4}^{\alpha \sigma }(x^{5})+\medskip \\
\\
-\delta ^{\alpha \rho }(\delta ^{\alpha 0}b_{0}^{-2}(x^{5})-\delta ^{\alpha
1}b_{1}^{-2}(x^{5})+\medskip \\
\\
-\delta ^{\alpha 2}b_{2}^{-2}(x^{5})-\delta ^{\alpha
3}b_{3}^{-2}(x^{5}))I_{DSR4}^{\beta \sigma }(x^{5})+\medskip \\
\\
-\delta ^{\beta \sigma }(\delta ^{\beta 0}b_{0}^{-2}(x^{5})-\delta ^{\beta
1}b_{1}^{-2}(x^{5})+\medskip \\
\\
-\delta ^{\beta 2}b_{2}^{-2}(x^{5})-\delta ^{\beta
3}b_{3}^{-2}(x^{5}))I_{DSR4}^{\alpha \rho }(x^{5})\medskip
\end{array}
\nonumber \\
\end{gather}
\begin{gather}
\text{4-d. deformed space-time translation algebra }tr.(3,1)_{DEF.}:\text{ }[%
\Upsilon _{DSR4}^{\mu }(x^{5}),\Upsilon _{DSR4}^{\nu }(x^{5})]=0  \nonumber
\\
\end{gather}
\begin{eqnarray}
&&
\begin{array}{c}
\text{4-d. {\em ''mixed''} deformed space-time} \\
\text{roto-translational algebra \medskip :}
\end{array}
\left\{
\begin{array}{l}
\lbrack I_{DSR4}^{\mu \nu }(x^{5}),\Upsilon _{DSR4}^{\rho }(x^{5})]=\medskip
\\
\\
\\
=g_{DSR4}^{\nu \rho }(x^{5})\Upsilon _{DSR4}^{\mu }-g_{DSR4}^{\mu \rho
}(x^{5})\Upsilon _{DSR4}^{\nu }= \\
\medskip \\
\\
\stackrel{\text{{\footnotesize ESC off}}}{=}\delta ^{\nu \rho
}(b_{0}^{-2}(x^{5})\delta ^{\nu 0}-b_{1}^{-2}(x^{5})\delta ^{\nu 1}+ \\
\medskip \\
-b_{2}^{-2}(x^{5})\delta ^{\nu 2}-b_{3}^{-2}(x^{5})\delta ^{\nu 3})\Upsilon
_{DSR4}^{\mu }(x^{5})+\medskip \\
\\
-\delta ^{\mu \rho }(b_{0}^{-2}(x^{5})\delta ^{\mu
0}-b_{1}^{-2}(x^{5})\delta ^{\mu 1}+\medskip \\
\\
-b_{2}^{-2}(x^{5})\delta ^{\mu 2}-b_{3}^{-2}(x^{5})\delta ^{\mu 3})\Upsilon
_{DSR4}^{\nu }(x^{5})\medskip ,
\end{array}
\right.  \nonumber \\
&&
\end{eqnarray}
or, in the ''self-representation'' deformed infinitesimal generator basis
\[
{\bf S}_{DSR4}(x^{5})\equiv
(I_{DSR4}^{23}(x^{5}),I_{DSR4}^{31}(x^{5}),I_{DSR4}^{12}(x^{5}))%
{\footnotesize \ }
\]
and
\[
{\bf K}_{DSR4}(x^{5})\equiv
(I_{DSR4}^{01}(x^{5}),I_{DSR4}^{02}(x^{5}),I_{DSR4}^{03}(x^{5})):
\]
\[
\]
\begin{gather}
\begin{array}{c}
\text{4-d. deformed space-time\medskip } \\
\text{rotation algebra} \\
\text{\ }su(2)_{DEF.}\times su(2)_{DEF.}\text{\medskip\ :}
\end{array}
\nonumber \\
\nonumber \\
\left\{
\begin{array}{l}
\lbrack S_{DSR4}^{i}(x^{5}),S_{DSR4}^{j}(x^{5})]\stackrel{\text{ESC off on }i%
\text{ and }j}{=}\medskip \\
\\
=\left( \sum_{s=1}^{3}(1-\delta _{is})((1-\delta
_{js})b_{s}^{-2}(x^{5})\right) \epsilon _{ijk}S_{DSR4}^{k}(x^{5})= \\
\\
=\medskip \medskip \epsilon _{ijk}b_{k}^{-2}(x^{5})S_{DSR4}^{k}(x^{5}) \\
\\
\\
\lbrack K_{DSR4}^{i}(x^{5}),K_{DSR4}^{j}(x^{5})]=\medskip \\
\\
=-b_{0}^{-2}(x^{5})\epsilon _{ijk}S_{DSR4}^{k}(x^{5})\medskip \medskip \\
\\
\\
\lbrack S_{DSR4}^{i}(x^{5}),K_{DSR4}^{j}(x^{5})]=\medskip \\
\\
\stackrel{\text{ESC off on }j}{=}\epsilon _{ijl}K_{DSR4}^{l}(x^{5})\left(
\sum_{s=1}^{3}\delta _{js}b_{s}^{-2}(x^{5})\right) \medskip = \\
\\
\stackrel{\text{ESC off on }j}{=}\epsilon
_{ijl}b_{j}^{-2}(x^{5})K_{DSR4}^{l}(x^{5})
\end{array}
\right.  \nonumber \\
\end{gather}
\[
\]
\begin{gather}
\text{4-d. deformed space-time translation algebra{\footnotesize \ }}%
tr.(3,1)_{DEF.}\text{{\footnotesize \ }:}  \nonumber \\
\nonumber \\
\lbrack \Upsilon _{DSR4}^{\mu }(x^{5}),\Upsilon _{DSR4}^{\nu
}(x^{5})]=0\bigskip
\end{gather}
\begin{gather}
\begin{array}{c}
\text{4-d. {\em ''mixed''} deformed space-time} \\
\text{roto-translational algebra\medskip\ :}
\end{array}
\nonumber \\
\nonumber \\
\nonumber \\
\left\{
\begin{array}{l}
\lbrack K_{DSR4}^{i}(x^{5}),\Upsilon
_{DSR4}^{0}(x^{5})]=-b_{0}^{-2}(x^{5})\Upsilon _{DSR4}^{i}(x^{5})\medskip \\
\\
\lbrack K_{DSR4}^{i}(x^{5}),\Upsilon _{DSR4}^{j}(x^{5})]\stackrel{\text{ESC
off on }i}{=}-\delta ^{ij}b_{i}^{-2}(x^{5})\Upsilon
_{DSR4}^{0}(x^{5})\medskip \\
\\
\lbrack S_{DSR4}^{i}(x^{5}),\Upsilon _{DSR4}^{0}(x^{5})]=0\medskip \\
\\
\lbrack S_{DSR4}^{i}(x^{5}),\Upsilon _{DSR4}^{k}(x^{5})]\stackrel{\text{ESC
off on }k}{=}\epsilon _{ikl}b_{k}^{-2}(x^{5})\Upsilon
_{DSR4}^{l}(x^{5}).\medskip
\end{array}
\right.  \nonumber \\
\end{gather}

\subsection{Explicit form of infinitesimal and finite deformed translations
in DSR4}

The $5\times 5$ matrix ${\cal T}_{T_{DSR4}^{\mu }(g,x^{5}),DSR4}(x^{5})$,
representing the infinitesimal (i.e. algebraic) element\footnote{%
For precision's sake, at the infinitesimal transformation level $\delta g\in
tr.(3,1)_{DEF.}\subset \left( \left( su(2)_{DEF.}\times su(2)_{DEF.}\right)
\times _{s}tr.(3,1)_{DEF.}\right) $ should be substituted for $g\in
Tr.(3,1)_{DEF.}\subset P(3,1)_{DEF.}$. But, for simplicity's sake, we will
omit, but understand, this cumbersome notation.} $\delta g\in
tr.(3,1)_{DEF.}\subset \left( \left( su(2)_{DEF.}\times su(2)_{DEF.}\right)
\times _{s}tr.(3,1)_{DEF.}\right) $\ which corresponds to a deformed,
infinitesimal 4-d. space-time translation by a parametric,
length-dimensioned (infinitesimal\footnote{%
Needless to say, at the algebraic and group level, ''length-dimensioned''
translation parameter contravariant deformed 4-vectors $T_{DSR4}^{\mu
}(g,x^{5})$ will be infinitesimal and finite, respectively. This will be
understood, and, for simplicity's sake, no notational distinction will be
made.
\par
As will be explicitly seen later (in the DSR4 case, without loss of
generality), the infinitesimal or finite nature of translation parameter $N$%
-vectors (such as $T_{DSR4}^{\mu }(g,x^{5})$ in DSR4) is in general the only
difference between algebraic and group level in translation coordinate
transformations in a generalized $N$-d. Minkowski space $\widetilde{M_{N}}%
\left( \left\{ x\right\} _{n.m.}\right) $.}) {\em contravariant} 4-vector
\begin{gather}
T_{DSR4}^{\mu }(g,x^{5})\equiv g_{DSR4}^{\mu \rho }(x^{5})T_{\rho
}(g)=\left(
b_{0}^{-2}(x^{5})T^{0},b_{1}^{-2}(x^{5})T^{1},b_{2}^{-2}(x^{5})T^{2},b_{3}^{-2}(x^{5})T^{3}\right) \medskip
\nonumber \\
\end{gather}
in $\widetilde{M_{4}}(x^{5})$, is defined by\footnote{%
The parentheses in the notation ''$DSR4$'' in $\Upsilon _{\mu ,\left(
DSR4\right) }$\ denote that, as expressed by Eqs. (11)-(12) and above
discussed - see also Footnote 10 -, the deformed translation infinitesimal
{\em covariant} generators are independent of the (geo)metric context being
considered.}
\begin{eqnarray}
{\cal T}_{T_{DSR4}^{\mu }(g,x^{5}),DSR4}(x^{5}) &\equiv &T_{DSR4}^{\mu
}(g,x^{5})\Upsilon _{\mu ,\left( DSR4\right) }=\medskip  \nonumber \\
&&  \nonumber \\
&=&g_{DSR4}^{\mu \rho }(x^{5})T_{\rho }(g)\Upsilon _{\mu ,\left( DSR4\right)
}=T_{\mu }(g)\Upsilon _{DSR4}^{\mu }(x^{5})=\medskip  \nonumber \\
&&  \nonumber \\
&=&T^{0}(g)\Upsilon _{DSR4}^{0}(x^{5})-T^{1}(g)\Upsilon _{DSR4}^{1}(x^{5})+
\nonumber \\
&&  \nonumber \\
&&-T^{2}(g)\Upsilon _{DSR4}^{2}(x^{5})-T^{3}(g)\Upsilon
_{DSR4}^{3}(x^{5})\medskip
\end{eqnarray}
or, explicitly:
\begin{equation}
{\cal T}_{T_{DSR4}^{\mu }(g,x^{5}),DSR4}(x^{5})=\left(
\begin{array}{ccccc}
0 & 0 & 0 & 0 & b_{0}^{-2}(x^{5})T^{0}(g) \\
0 & 0 & 0 & 0 & b_{1}^{-2}(x^{5})T^{1}(g) \\
0 & 0 & 0 & 0 & b_{2}^{-2}(x^{5})T^{2}(g) \\
0 & 0 & 0 & 0 & b_{3}^{-2}(x^{5})T^{3}(g) \\
0 & 0 & 0 & 0 & 0
\end{array}
\right) .
\end{equation}

Therefore, the 4-d. deformed, infinitesimal space-time translation by
(infinitesimal)
\[
T_{DSR4}^{\mu }(g,x^{5})\stackrel{\text{ESC on}}{\equiv }g_{DSR4}^{\mu \rho
}(x^{5})T_{\rho }(g)
\]
in $\widetilde{M_{4}}(x^{5})$ - corresponding to
\[
\delta g\in tr.(3,1)_{DEF.}\subset \left( \left( su(2)_{DEF.}\times
su(2)_{DEF.}\right) \times _{s}tr.(3,1)_{DEF.}\right) -
\]
is given by (ESC on; for simplicity's sake, dependence on $\left\{
x_{m.}\right\} $ is omitted):
\begin{gather}
\left(
\begin{array}{c}
\left( x^{\prime }\right) _{(g),DSR4}^{0}(x^{0},x^{5}) \\
\left( x^{\prime }\right) _{(g),DSR4}^{1}(x^{1},x^{5}) \\
\left( x^{\prime }\right) _{(g),DSR4}^{2}(x^{2},x^{5}) \\
\left( x^{\prime }\right) _{(g),DSR4}^{3}(x^{3},x^{5}) \\
y_{DSR4}^{\prime }
\end{array}
\right) =\left(
\begin{array}{c}
\left( x^{0}\right) _{(g),DSR4}^{^{\prime }}(x^{0},x^{5}) \\
\left( x^{1}\right) _{(g),DSR4}^{^{\prime }}(x^{1},x^{5}) \\
\left( x^{2}\right) _{(g),DSR4}^{^{\prime }}(x^{2},x^{5}) \\
\left( x^{3}\right) _{(g),DSR4}^{^{\prime }}(x^{3},x^{5}) \\
y_{DSR4}^{\prime }
\end{array}
\right) =  \nonumber \\
\nonumber \\
\stackrel{\text{ESC on}}{=}\left( 1_{5-d.}+{\cal T}_{T_{DSR4}^{\mu
}(g,x^{5}),DSR4}(x^{5})\right) _{~~B}^{A}x^{B}=\bigskip  \nonumber \\
\nonumber \\
=\left(
\begin{array}{ccccc}
1 & 0 & 0 & 0 & b_{0}^{-2}(x^{5})T^{0}(g) \\
0 & 1 & 0 & 0 & b_{1}^{-2}(x^{5})T^{1}(g) \\
0 & 0 & 1 & 0 & b_{2}^{-2}(x^{5})T^{2}(g) \\
0 & 0 & 0 & 1 & b_{3}^{-2}(x^{5})T^{3}(g) \\
0 & 0 & 0 & 0 & 1
\end{array}
\right) \left(
\begin{array}{c}
x^{0} \\
x^{1} \\
x^{2} \\
x^{3} \\
\left( y=\right) 1
\end{array}
\right) =  \nonumber \\
\nonumber \\
=\left(
\begin{array}{c}
x^{0}+b_{0}^{-2}(x^{5})T^{0}(g) \\
x^{1}+b_{1}^{-2}(x^{5})T^{1}(g) \\
x^{2}+b_{2}^{-2}(x^{5})T^{2}(g) \\
x^{3}+b_{3}^{-2}(x^{5})T^{3}(g) \\
\left( y=\right) 1
\end{array}
\right) \bigskip .
\end{gather}

At the finite transformation level, one has to evaluate the exponential of
the matrix ${\cal T}_{T_{DSR4}^{\mu }(g,x^{5}),DSR4}(x^{5})$, i.e. the $%
5\times 5$ matrix $\exp \left( {\cal T}_{T_{DSR4}^{\mu
}(g,x^{5}),DSR4}(x^{5})\right) $, representing the finite (i.e. group)
element $g\in $ $Tr.(3,1)_{DEF.}\subset SO(3,1)_{DEF.}$\ which corresponds
to a deformed, finite 4-d. space-time translation by a parametric,
length-dimensioned (finite) {\em contravariant} 4-vector $T_{DSR4}^{\mu
}(g,x^{5})\stackrel{\text{ESC on}}{\equiv }g_{DSR4}^{\mu \rho
}(x^{5})T_{\rho }(g)$ :
\begin{eqnarray}
\exp \left( {\cal T}_{T_{DSR4}^{\mu }(g,x^{5}),DSR4}(x^{5})\right)
&=&\sum_{n=0}^{\infty }\frac{1}{n!}\left( {\cal T}_{T_{DSR4}^{\mu
}(g,x^{5}),DSR4}(x^{5})\right) ^{n}=  \nonumber \\
&&  \nonumber \\
&=&1_{5-d.}+{\cal T}_{T_{DSR4}^{\mu }(g,x^{5}),DSR4}(x^{5})\medskip . \\
&&  \nonumber
\end{eqnarray}

Then, as anticipated in Footnote 12, for translation transformations the
only difference between the infinitesimal level and the finite one is
provided by the infinitesimal or finite nature of the {\em contravariant}
translation parameters \ $T_{DSR4}^{\mu }(g,x^{5})$. Such a result is a
peculiar feature of the space-time translation component of the 4-d.
deformed Poincar\'{e} group $P(3,1)_{DEF.}$, and in general of translation
coordinate transformations in $N$-d. generalized Minkowski spaces $%
\widetilde{M_{N}}\left( \left\{ x\right\} _{n.m.}\right) $. Still
considering the case of DSR4, it can be recovered also by exploiting the
Abelian nature of $Tr.(3,1)_{DEF.}$ and the property of the powers of its
infinitesimal generators (see Eq. (25)), and by using the {\em %
Baker-Campbell-Hausdorff formula} [11]; one has indeed:
\begin{gather}
\exp \left( {\cal T}_{T_{DSR4}^{\mu }(g,x^{5}),DSR4}(x^{5})\right) =\medskip
\nonumber \\
\nonumber \\
=\exp (T^{0}(g)\Upsilon _{DSR4}^{0}(x^{5})-T^{1}(g)\Upsilon
_{DSR4}^{1}(x^{5})+\medskip  \nonumber \\
\nonumber \\
-T^{2}(g)\Upsilon _{DSR4}^{2}(x^{5})-T^{3}(g)\Upsilon
_{DSR4}^{3}(x^{5}))=\medskip  \nonumber \\
\nonumber \\
=\exp \left( T^{0}(g)\Upsilon _{DSR4}^{0}(x^{5})\right) \times \medskip
\nonumber \\
\nonumber \\
\times \exp \left( T^{1}(g)\Upsilon _{DSR4}^{1}(x^{5})\right) \times \exp
\left( T^{2}(g)\Upsilon _{DSR4}^{2}(x^{5})\right) \times \exp \left(
T^{3}(g)\Upsilon _{DSR4}^{3}(x^{5})\right) =  \nonumber \\
\medskip  \nonumber \\
=\left( \sum_{n=0}^{\infty }\frac{\left( T^{0}(g)\right) ^{n}}{n!}\left(
\Upsilon _{DSR4}^{0}(x^{5})\right) ^{n}\right) \times \left(
\sum_{n=0}^{\infty }\frac{\left( T^{1}(g)\right) ^{n}}{n!}\left( \Upsilon
_{DSR4}^{1}(x^{5})\right) ^{n}\right) \times \medskip  \nonumber \\
\nonumber \\
\times \left( \sum_{n=0}^{\infty }\frac{\left( T^{2}(g)\right) ^{n}}{n!}%
\left( \Upsilon _{DSR4}^{2}(x^{5})\right) ^{n}\right) \times \left(
\sum_{n=0}^{\infty }\frac{\left( T^{3}(g)\right) ^{n}}{n!}\left( \Upsilon
_{DSR4}^{3}(x^{5})\right) ^{n}\right) =\medskip  \nonumber \\
\nonumber \\
=\left( 1_{5-d.}+T^{0}(g)\Upsilon _{DSR4}^{0}(x^{5})\right) \times \left(
1_{5-d.}+T^{1}(g)\Upsilon _{DSR4}^{1}(x^{5})\right) \times \medskip
\nonumber \\
\nonumber \\
\times \left( 1_{5-d.}+T^{2}(g)\Upsilon _{DSR4}^{2}(x^{5})\right) \times
\left( 1_{5-d.}+T^{3}(g)\Upsilon _{DSR4}^{3}(x^{5})\right) =\medskip
\nonumber \\
\nonumber \\
=1_{5-d.}+T^{0}(g)\Upsilon _{DSR4}^{0}(x^{5})+T^{1}(g)\Upsilon
_{DSR4}^{1}(x^{5})+T^{2}(g)\Upsilon _{DSR4}^{2}(x^{5})+\medskip  \nonumber \\
\nonumber \\
+T^{3}(g)\Upsilon _{DSR4}^{3}(x^{5})=  \nonumber \\
\nonumber \\
\nonumber \\
1_{5-d.}+{\cal T}_{T_{DSR4}^{\mu }(g,x^{5}),DSR4}(x^{5})\medskip .
\end{gather}

On account of the non-commutativity of the infinitesimal generators of $%
Tr.(3,1)_{DEF.}$ and of $SO(3,1)_{DEF.}$ (see Eqs. (36) and (37)), by using
the {\em Baker-Campbell-Hausdorff formula} [11], it is possible to state the
following unequality for the considered 5-d. matrix representing the finite
(i.e. group) element $g$ $\in $ $P(3,1)_{DEF.}$ corresponding to a finite,
4-d. deformed space-time roto-translation in $\widetilde{M_{4}}(x^{5})$, of
dimensionless parametric deformed angular (Euclidean) 3-vector ${\bf \theta }%
(g)$, dimensionless parametric deformed ''rapidity'' (Euclidean) 3-vector $%
{\bf \zeta }(g)$ and ''length-dimensioned'' parametric translational
contravariant (deformed) 4-vector $T_{DSR4}^{\mu }(g,x^{5})\equiv $ $%
g_{DSR4}^{\mu \rho }(x^{5})T_{\rho }(g)$ \footnote{%
Let us instead notice that, at infinitesimal level, all transformations of
the Lie group $P(3,1)_{DEF.}$ commute.}:
\[
\]
\begin{gather}
\exp \left( -{\bf \theta }(g)\cdot {\bf S}_{DSR4}(x^{5})-{\bf \zeta }%
(g)\cdot {\bf K}_{DSR4}(x^{5})+T_{\mu }(g)\Upsilon _{DSR4}^{\mu
}(x^{5})\right) \neq \medskip  \nonumber \\
\nonumber \\
\neq \exp \left( -{\bf \theta }(g)\cdot {\bf S}_{DSR4}(x^{5})-{\bf \zeta }%
(g)\cdot {\bf K}_{DSR4}(x^{5})\right) \times \exp \left( T_{\mu }(g)\Upsilon
_{DSR4}^{\mu }(x^{5})\right) =\medskip  \nonumber \\
\nonumber \\
=\exp \left( -{\bf \theta }(g)\cdot {\bf S}_{DSR4}(x^{5})-{\bf \zeta }%
(g)\cdot {\bf K}_{DSR4}(x^{5})\right) \times \exp \left( {\cal T}%
_{T_{DSR4}^{\mu }(g,x^{5}),DSR4}(x^{5})\right) =\medskip  \nonumber \\
\nonumber \\
=\exp \left( -{\bf \theta }(g)\cdot {\bf S}_{DSR4}(x^{5})-{\bf \zeta }%
(g)\cdot {\bf K}_{DSR4}(x^{5})\right) \times \left( 1_{5-d.}+{\cal T}%
_{T_{DSR4}^{\mu }(g,x^{5}),DSR4}(x^{5})\right) \neq \medskip  \nonumber \\
\nonumber \\
\neq \exp \left( -{\bf \theta }(g)\cdot {\bf S}_{DSR4}(x^{5})\right) \times
\exp \left( -{\bf \zeta }(g)\cdot {\bf K}_{DSR4}(x^{5})\right) \times \left(
1_{5-d.}+{\cal T}_{T_{DSR4}^{\mu }(g,x^{5}),DSR4}(x^{5})\right) \neq
\nonumber \\
\nonumber \\
\neq \medskip \exp \left( -\theta ^{1}(g)S_{DSR4}^{1}(x^{5})\right) \times
\exp \left( -\theta ^{2}(g)S_{DSR4}^{2}(x^{5})\right) \times  \nonumber \\
\nonumber \\
\times \exp \left( -\theta ^{3}(g)S_{DSR4}^{3}(x^{5})\right) \times \exp
\left( -\zeta ^{1}(g)K_{DSR4}^{1}(x^{5})\right) \times  \nonumber \\
\nonumber \\
\times \exp \left( -\zeta ^{2}(g)K_{DSR4}^{2}(x^{5})\right) \times \exp
\left( -\zeta ^{3}(g)K_{DSR4}^{3}(x^{5})\right) \times \left( 1_{5-d.}+{\cal %
T}_{T_{DSR4}^{\mu }(g,x^{5}),DSR4}(x^{5})\right) ,  \nonumber \\
\end{gather}

where in the last two lines the non commutativity of deformed true rotation
and ''boost'' infinitesimal generators have been used (see Eqs. (41) and
(44)).

By comparing Eq. (50) with the expression of a translation in the usual
Minkowski space $M_{4}$ of SR (as before ESC on and, for simplicity's sake,
dependence on $\left\{ x_{m.}\right\} $ is omitted)
\begin{gather}
\left(
\begin{array}{c}
\left( x^{\prime }\right) _{(g),SSR4}^{0}(x^{0}) \\
\left( x^{\prime }\right) _{(g),SSR4}^{1}(x^{1}) \\
\left( x^{\prime }\right) _{(g),SSR4}^{2}(x^{2}) \\
\left( x^{\prime }\right) _{(g),SSR4}^{3}(x^{3}) \\
y_{SSR4}^{\prime }
\end{array}
\right) =\left(
\begin{array}{c}
\left( x^{0}\right) _{(g),SSR4}^{^{\prime }}(x^{0}) \\
\left( x^{1}\right) _{(g),SSR4}^{^{\prime }}(x^{1}) \\
\left( x^{2}\right) _{(g),SSR4}^{^{\prime }}(x^{2}) \\
\left( x^{3}\right) _{(g),SSR4}^{^{\prime }}(x^{3}) \\
y_{SSR4}^{\prime }
\end{array}
\right) =  \nonumber \\
\nonumber \\
=\left( 1_{5-d.}+{\cal T}_{T_{SSR4}^{\mu }(g),SSR4}\right)
_{~~B}^{A}x^{B}=\bigskip  \nonumber \\
\nonumber \\
=\left(
\begin{array}{ccccc}
1 & 0 & 0 & 0 & T^{0}(g) \\
0 & 1 & 0 & 0 & T^{1}(g) \\
0 & 0 & 1 & 0 & T^{3}(g) \\
0 & 0 & 0 & 1 & T^{3}(g) \\
0 & 0 & 0 & 0 & 1
\end{array}
\right) \left(
\begin{array}{c}
x^{0} \\
x^{1} \\
x^{2} \\
x^{3} \\
\left( y=\right) 1
\end{array}
\right) =\left(
\begin{array}{c}
x^{0}+T^{0}(g) \\
x^{1}+T^{1}(g) \\
x^{2}+T^{2}(g) \\
x^{3}+T^{3}(g) \\
\left( y=\right) 1
\end{array}
\right) \bigskip , \\
\nonumber \\
\nonumber
\end{gather}
it is easily seen that passing from SR to DSR4 - i.e. {\em locally deforming
and spatially anisotropizing} $M_{4}$ - (as far as space-time translations
are concerned) amounts to the following parameter change:
\begin{eqnarray}
T_{SSR4}^{\mu }(g) &=&\left( T^{0}(g),T^{1}(g),T^{2}(g),T^{3}(g)\right)
\rightarrow T_{DSR4}^{\mu }(g,x^{5})=\medskip  \nonumber \\
&&  \nonumber \\
&=&\left(
b_{0}^{-2}(x^{5})T^{0}(g),b_{0}^{-2}(x^{5})T^{1}(g),b_{0}^{-2}(x^{5})T^{2}(g),b_{0}^{-2}(x^{5})T^{3}(g)\right) \equiv
\widetilde{T}_{DSR4}^{\mu }(g,x^{5})\medskip .  \nonumber \\
&&
\end{eqnarray}

Then, extending the meaning of ''{\em effective}'' transformation parameters
[2] to translation ones, we can say that in the translational case the
''length-dimensioned'' deformed translation parameter (deformed) {\em %
contravariant} 4-vector $T_{DSR4}^{\mu }(g,x^{5})$ coincides with the {\em %
effective} ''length-dimensioned'' deformed translation parameter (deformed)
{\em contravariant }4-vector\footnote{%
The contravariant 4-vector identity (56) is due to the very fact that the
passage SR$\rightarrow $DSR4:
\[
g_{\mu \nu ,SR}=diag\left( 1,-1,-1,-1\right) \rightarrow g_{\mu \nu
,DSR4}(x^{5})=diag\left(
b_{0}^{2}(x^{5}),-b_{1}^{2}(x^{5}),-b_{2}^{2}(x^{5}),-b_{3}^{2}(x^{5})%
\right)
\]
{\em preserves} the diagonality of the 2-rank, symmetric metric 4-tensor,
still {\em destroying} its isochrony and spatial isotropy.} $\widetilde{T}%
_{DSR4}^{\mu }(g,x^{5})\medskip $ :
\begin{equation}
T_{DSR4}^{\mu }(g,x^{5})=\widetilde{T}_{DSR4}^{\mu }(g,x^{5})\medskip .
\end{equation}
This is peculiar feature of the translation component $Tr.(3,1)_{DEF.}$ of
the 4-d. deformed Poincar\`{e} group $P(3,1)_{DEF.}$, and it is due to the
following fact: while for 4-d. space-time rotations (homogeneous
transformations in the coordinates) the deformed transformation parameter
3-vectors ${\bf \theta }\left( g\right) $ and ${\bf \zeta }\left( g\right) $
are Euclidean (see Refs. [1] and [2]), in the case of 4-d. deformed
translations (inhomogeneous transformations in the coordinates) the deformed
translation parameter {\em contravariant} 4-vector $T_{DSR4}^{\mu }(g,x^{5})$
is ''length-dimensioned'' and deformed, i.e. {\em dependent on the metric
structure being considered} (see also Footnotes 5, 10 and 13).

Let us finally stress that the {\em (local) ''deformating anisotropizing''}
generalization of SR corresponding to DSR4 is fully self-consistent at
space-time translation level, too. As noticed also in Refs. [1], [2] and
[12], it is easy to see that {\em all} the results obtained in the present
work for the DSR4 level (i.e. about the Lie group $Tr.(3,1)_{DEF.}$ of
space-time deformed translations in the 4-d. ''deformed'' Minkowski space $%
\widetilde{M_{4}}(x^{5})$) reduce, in the limit DSR4$\rightarrow $SR, i.e.
in the limit
\begin{eqnarray}
g_{\mu \nu ,DSR4}(x^{5}) &=&diag\left(
b_{0}^{2}(x^{5}),-b_{1}^{2}(x^{5}),-b_{2}^{2}(x^{5}),-b_{3}^{2}(x^{5})%
\right) \rightarrow  \nonumber \\
&&  \nonumber \\
&\rightarrow &g_{\mu \nu ,SR}=diag\left( 1,-1,-1,-1\right) \Leftrightarrow
\medskip  \nonumber \\
&&  \nonumber \\
&\Leftrightarrow &b_{\mu }^{2}(x^{5})\rightarrow 1,\forall \mu
=0,1,2,3\medskip ,
\end{eqnarray}
to the well-known results of SR (i.e. about the Lie group $Tr.(3,1)_{STD.}$
of space-time translations in the 4-d. Minkowski space $M_{4}$ of Einstein's
Special Relativity).

\end{document}